\documentclass[12pt,a4]{article} 
\newcommand{\vs}{\vspace{-0.25cm}}
\usepackage{amsmath,graphicx}
\begin{document} 
\begin{center}
  {\Large{\bf Density-dependent nn-potential from subleading chiral
three-neutron forces: Long-range terms}\footnote{This work 
has been supported in part by DFG and NSFC (CRC110).}  }  

\medskip

 N. Kaiser \\
\medskip
{\small Physik-Department T39, Technische Universit\"{a}t M\"{u}nchen,
   D-85747 Garching, Germany}
\end{center}
\medskip
\begin{abstract}
The long-range terms of the subleading chiral three-nucleon force [published in Phys.\,Rev.\,C77, 064004 (2008)] are specified to the case of three neutrons. From these $3n$-interactions an effective density-dependent neutron-neutron potential $V_\text{med}$ in pure neutron matter is derived. 
Following the division of the pertinent 3n-diagrams into two-pion exchange, two-pion-one-pion exchange and ring topology, all self-closings and concatenations of two neutron-lines to an in-medium loop are evaluated. The momentum and $k_n$-dependent potentials associated with the  spin-operators $1,\, \vec\sigma_1\!\cdot\!\vec\sigma_2,\, \vec\sigma_1\!\cdot\!\vec q\, \vec\sigma_2\!\cdot\!\vec q,\, i( \vec\sigma_1\!+\!\vec\sigma_2)\!\cdot \! (\vec q\!\times \! \vec p\,),\, (\vec\sigma_1\!\cdot\!\vec p\,\vec\sigma_2\!\cdot\!\vec p+\vec\sigma_1\!\cdot\!\vec p\,'\, \vec\sigma_2\!\cdot\!\vec p\,')$ and $ \vec\sigma_1\!\cdot \! (\vec q\!\times \! \vec p\,)\vec\sigma_2\!\cdot \! (\vec q\!\times \! \vec p\,)$ are expressed in terms of functions, which are either given in closed analytical form or require at most one numerical integration. The subsubleading chiral 3N-force is treated in the same way. The obtained results for $V_\text{med}$ are helpful to implement the long-range chiral three-body forces into advanced neutron matter  calculations.
\end{abstract}

\section{Introduction}
Three-nucleon forces are an indispensable ingredient in accurate few-nucleon and
nuclear structure calculations. Nowadays, chiral effective field theory is the
appropriate tool to construct systematically the nuclear interactions in harmony with the symmetries of QCD. Three-nucleon forces appear first at N$^2$LO, where they consist of a zero-range contact-term ($\sim c_E$) , a mid-range $1\pi$-exchange component ($\sim c_D$) and a long-range $2\pi$-exchange component ($\sim c_{1,3,4}$). The complete calculation of the chiral 3N-forces to subleading order 
N$^3$LO \cite{3Nlong,3Nshort} and even to subsubleading order N$^4$LO \cite{twopi4,midrange4} has been achieved during the past decade by the Bochum-Bonn group. At present the focus lies on constructing 3N-forces in chiral effective field theory with explicit $\Delta(1232)$-isobars, for which the long-range $2\pi$-exchange component  has been derived recently in ref.\,\cite{twopidelta} at order N$^3$LO.

However, for the variety of existing many-body methods, that are commonly employed in calculations of nuclear matter or medium mass and heavy nuclei, it is technically very challenging to include the chiral three-nucleon forces directly. An alternative and approximate approach is to use instead a density-dependent two-nucleon interaction $V_\text{med}$ that originates from the underlying 3N-force. When restricting to on-shell scattering of two nucleons in isospin-symmetric spin-saturated nuclear matter, $N_1(\vec p\,)+N_2(-\vec p\,)\to N_1(\vec p\,')+N_2(-\vec p\,')$, the resulting in-medium NN-potential  $V_\text{med}$ has the same isospin- and spin-structure as the free NN-potential. The analytical expressions for $V_\text{med}$ from the leading chiral 3N-force at 
N$^2$LO (involving the parameters $c_{1,3,4}$, $c_D$ and $c_E$) have been presented in ref.\,\cite{holt} and these have found many applications in recent years.  But in order to perform nuclear many-body calculations that are consistent with their input at the two-body level, one needs also $V_\text{med}$ derived from the subleading chiral 3N-forces at order N$^3$LO. In two recent works this task has been completed for the short-range terms and relativistic $1/M$-corrections in ref.\,\cite {vmedshort}, and for the long-range terms in ref.\,\cite{vmedlong}. The continuation of this construction from the intermediate-range terms of the subsubleading chiral 3N-force at order N$^3$LO has been reported recently in ref.\cite{subsubvmed}. The normal ordering of chiral 3N-forces to density-dependent NN-interactions has also been performed by the Darmstadt group using a decomposition in a $Jj$-coupled 3N partial-wave momentum basis \cite{normalorder,achim1}. An advantage of this purely numerical approach is that the restriction to on-shell kinematics in the center-of-mass frame can be avoided and a regulator function $f_{R}(p,q)=\exp[-(p^2+3q^2/4)^4/\Lambda^8_{\rm 3N}]$ may be included.

In this work  the specification to the case of three neutrons is considered and an effective density-dependent neutron-neutron potential $V_\text{med}$ in pure neutron matter is derived.  
The 3n-interaction is obtained from a given expression for the 3N-force by the simple substitution of isospin-operators: $\vec\tau_1 \!\cdot\! \vec\tau_2\to 1, \vec\tau_1 \!\cdot\! \vec\tau_3\to1, \vec\tau_2 \!\cdot\! \vec\tau_3\to1$ and $\vec\tau_1 \!\cdot\!(\vec\tau_2\!\times\!\vec\tau_3)\to 0$. In the next step three  self-closings and six concatenations of two neutron-lines to an in-medium loop need to be evaluated. This introduces a set of loop-functions that depend on the momentum $p=| \vec p\,|$, the momentum-transfer $ q= |\vec p\,'-\vec p\,|$, and the neutron Fermi-momentum $k_n$. Since the calculational procedure follows closely that in previous works \cite{vmedshort,vmedlong,subsubvmed}, it is sufficient to list the results for the nn-potential $V_\text{med}$ in pure neutron matter without further explanation. The analytical results for the in-medium nn-potential $V_\text{med}$ derived from the subleading short-range terms and relativistic $1/M$-corrections have been reported in ref.\cite{treuer}. 
\begin{figure}[h]
\centering
\includegraphics[width=0.8\textwidth]{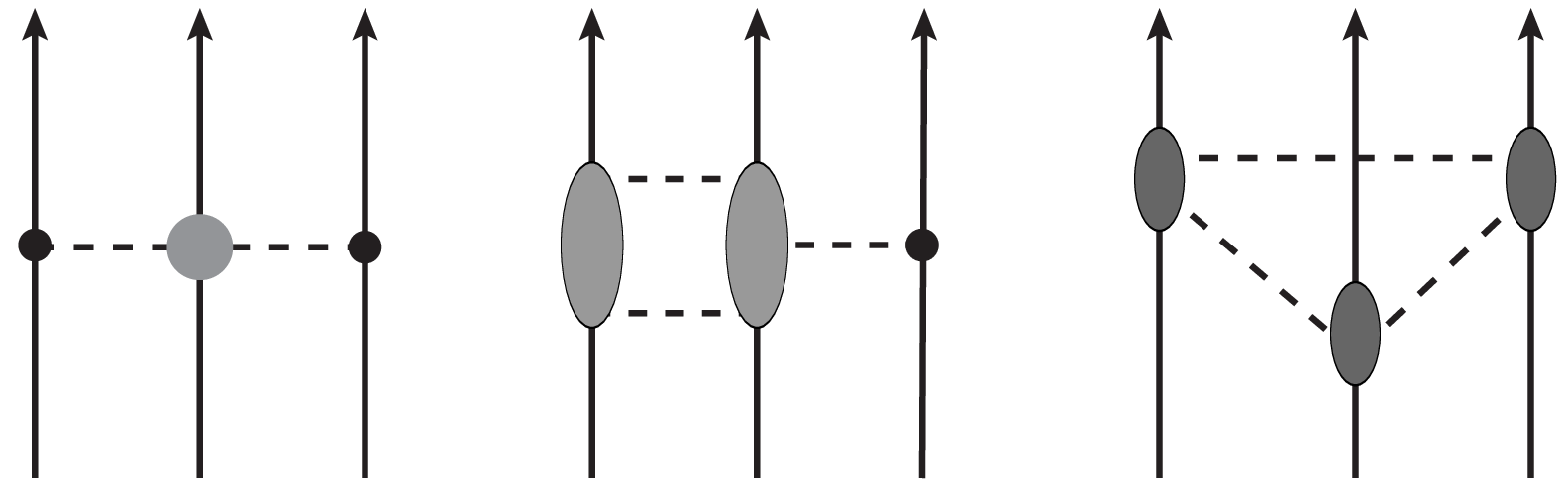}
\caption{$2\pi$-exchange topology, $2\pi1\pi$-exchange topology and ring topology which comprise the long- and  intermediate-range chiral 3n-forces.} \end{figure}
\section{Two-pion exchange topology}
One starts with the longest-range component of the subleading chiral 3N-interaction. It arises from $2\pi$-exchange with the corresponding symbolic diagram shown on the left of Fig.\,1. Using for simplification the relation $q_2^2 = q_1^2+q_3^2 +2\vec q_1 \!\cdot\!\vec q_3$, the expression in eq.(2.9) of ref.\,\cite{3Nlong} specified to three neutrons takes the form:
\begin{eqnarray}
2V_\text{3n}&=&{g_A^4 \over 128\pi f_\pi^6}\, {\vec\sigma_1\!\cdot\! \vec q_1 \vec
\sigma_3\!\cdot\! \vec q_3 \over (m_\pi^2+q_1^2) (m_\pi^2+q_3^2)}\Big\{m_\pi(m_\pi^2+q_1^2+q_3^2+2q_2^2) \nonumber \\ && +
(2m_\pi^2+q_2^2)(3m_\pi^2+q_1^2+q_3^2+2q_2^2)A(q_2) \Big\} \,, \end{eqnarray}
  with the pion-loop function $A(s)= (1/2s) \arctan(s/2m_\pi)$.
\subsection{Contributions to in-medium nn-potential}
The self-closing of neutron-line $2$ gives (after relabeling $3\to 2$) the contribution
\begin{equation}V^{(0)}_\text{med} = - {g^4_A m_\pi k^3_n \over 2(4\pi f^2_\pi)^3}\,
{\vec\sigma_1\!\cdot\!\vec q\, \vec\sigma_2\!\cdot\!\vec q\over (m^2_\pi+q^2)^2} \Big(q^2+{5m^2_\pi\over 6}\Big)\,. \end{equation}
From pionic vertex corrections on either neutron-line one obtains the (total) contribution:
\begin{eqnarray}V^{(1)}_\text{med} &=&  {2g^4_A \over (8\pi f^2_\pi)^3}\,
{\vec\sigma_1\!\cdot\!\vec q\, \vec\sigma_2\!\cdot\!\vec q\over m^2_\pi+q^2}\Big\{m_\pi\big[ 2k_n^3 -8\Gamma_2-2m_\pi^2 (\Gamma_0+\Gamma_1) \nonumber \\  &&  +q^2 (\Gamma_0-\Gamma_1-2\Gamma_3)\big]  +
S(p) + J(p,q)\Big\}\,,\end{eqnarray}
where the functions $\Gamma_\nu(p,k_n)$ are defined in the appendix of
ref.\,\cite{vmedshort}. The part linear in $m_\pi$ comes obviously from the first line in eq.(1) and the decomposition $S(p) + J(p,q)$ is obtained by canceling momentum factors against a pion-propagator. The two functions appearing at the end of eq.(3) read:
\begin{eqnarray}S(p)&\!\!\!=\!\!\!& \bigg[ {k_n^5\over 15p^3}\Big(m_\pi^2+{k_n^2\over 14}\Big)+{k_n^3m_\pi^2 \over 3p}+{k_n^2\over 3}\Big({k_n^2\over 4}+2m_\pi^2\Big)+{k_n^3p \over 6}+{p^2\over5}\Big({k_n^2\over 2}-{4m_\pi^2\over3}\Big) -{p^4\over84} \bigg] \arctan{p+ k_n \over 2m_\pi} \nonumber \\ && + \bigg[ {k_n^5\over 15p^3}\Big(m_\pi^2\!+\!{k_n^2\over 14}\Big)+{k_n^3m_\pi^2 \over 3p}-{k_n^2\over 3}\Big({k_n^2\over 4}+2m_\pi^2\Big)+{k_n^3p \over 6}+{p^2\over5}\Big({4m_\pi^2\over3}-{k_n^2\over 2}\Big) +{p^4\over84} \bigg] \arctan{p- k_n \over 2m_\pi}\nonumber \\ && +m_\pi^3 \bigg[ {7p\over 24}+{1\over p}\Big({m_\pi^2 \over 3}-{k_n^2 \over 4}\Big)+{1 \over p^3}\Big({6m_\pi^4 \over 35}+{k_n^2 m_\pi^2 \over 15}-{k_n^4 \over 24}\Big) \bigg] \ln{4m_\pi^2+(p+k_n)^2\over 4m_\pi^2+(p-k_n)^2}\nonumber \\ && +{k_n m_\pi \over 7}\bigg[{p^2\over 3}- {61m_\pi^2\over 30}-{37k_n^2\over 15}-{1\over p^2}\Big({2k_n^4\over 15}+{k_n^2m_\pi^2\over 6}+{6m_\pi^4\over 5}\Big)\bigg]\,, \end{eqnarray}
\begin{eqnarray}J(p,q) &=& {1\over 4q^2} \int_{p-k_n}^{p+k_n}\!ds
(2m_\pi^2+s^2)(2m_\pi^2+q^2 +2s^2) \arctan{ s\over 2m_\pi}\nonumber \\ && \times
\bigg\{{k_n^2 -(p-s)^2 \over p} + {q^2-m_\pi^2-s^2 \over q} \ln{ q X + 2\sqrt{W}
\over (2p+q)[m_\pi^2+(q-s)^2]}\bigg\} \,,  \end{eqnarray}
with the auxiliary polynomials:
\begin{eqnarray} && X = m_\pi^2 +2(k_n^2-p^2)+q^2-s^2\,, \nonumber \\
&& W = k_n^2 q^4+p^2(m_\pi^2+s^2)^2 +q^2\big[ (k_n^2-p^2)^2+m_\pi^2(k_n^2+p^2)
   -s^2(k_n^2+p^2+m_\pi^2)\big]\,.  \end{eqnarray}
 Finally, the two diagrams related to double exchange lead to the expression:
 \begin{eqnarray} V^{(3)}_\text{med} & = &{g^4_A\over (8\pi  f^2_\pi)^3} \bigg\{
\big[ m_\pi+(2m_\pi^2+q^2) A(q) \big]\Big[{8k_n^3\over3}+2q^2(\Gamma_0-\Gamma_1)
\Big] -2m_\pi^2\Gamma_0\big[3m_\pi\nonumber\\ && + (2m_\pi^2+q^2) A(q)\big]+
(2m_\pi^2+q^2)\big[m_\pi^3-2m_\pi q^2-(2m_\pi^4+5m_\pi^2q^2+2q^4)A(q)\big] G_{0}
 \nonumber\\ && + i( \vec\sigma_1\!+\!\vec\sigma_2)\!\cdot\!(\vec q\!\times \!
 \vec p\,)\Big\{ 2\big[m_\pi+(2m_\pi^2+q^2)A(q) \big](\Gamma_0+ \Gamma_1)\nonumber\\ &&+\big[2m_\pi q^2-m_\pi^3+(2m_\pi^4+5m_\pi^2q^2+2q^4)A(q)\big](G_0+2G_1)
\Big\}\bigg\}\,.  \end{eqnarray}

\section{Two-pion-one-pion exchange topology}
Next, one considers the $2\pi1\pi$-exchange three-neutron interaction represented by the middle diagram in Fig.\,1. According to eqs.(2.16)-(2.20) in ref.\cite{3Nlong} this chiral 3n-interaction  can be written in the form:
\begin{eqnarray}
V_\text{3n}&=&{g_A^4 \over 256\pi f_\pi^6} { \vec\sigma_3\!\cdot\! \vec q_3 \over m_\pi^2+q_3^2}\Big\{\vec\sigma_2\!\cdot\! \vec q_1\,\vec q_1\!\cdot\! \vec q_3\, f_1(q_1) + \vec\sigma_2\!\cdot\! \vec q_1\,f_2(q_1)\nonumber\\ && + \vec\sigma_2\!\cdot\! \vec q_3\,f_3(q_1)+\vec\sigma_1\!\cdot\! \vec q_1\,\vec q_1\!\cdot\! \vec q_3\, f_4(q_1) + \vec\sigma_1\!\cdot\! \vec q_3\,f_5(q_1)+\vec\sigma_2\!\cdot\! \vec q_3\,\vec q_1\!\cdot\! \vec q_3\, f_8(q_1) \Big\}\,,  \end{eqnarray}
where a common factor $g_A^4/(256\pi f_\pi^6)$ has been pulled out, and thus the reduced  functions $f_j(s)$ read: 
\begin{eqnarray}&& f_1(s) = {m_\pi\over s^2}(1-2 g_A^2)-{g_A^2m_\pi\over 4m_\pi^2 + s^2}  +\Big[1+g_A^2 +{4 m_\pi^2 \over s^2}(2 g_A^2-1)\Big]  A(s)  \,,
\\ && f_2(s)= 6m_\pi+6(2m_\pi^2+s^2)A(s) \,,
\\ && f_3(s)= (1-g_A^2)\big[3m_\pi+(8m_\pi^2+3s^2)A(s)\big] \,, \\  &&  f_5(s)=-s^2 f_4(s)=2 g_A^2 s^2 A(s)\,,  \end{eqnarray}
with $A(s)$ defined after eq.(1). In perspective one notes that at subsubleading order \cite{midrange4} the chiral $2\pi1\pi$-exchange 3n-interaction has a richer spin- and momentum-dependence and also the last term involving  $f_8(q_1)$ is needed to represent all diagrams belonging to this topology at N$^4$LO. In the 
three-neutron case the other relevant functions are combined as: $f_1(s) \equiv f_1(s)+f_6(s),\, f_2(s) \equiv f_2(s)+f_7(s),\, f_3(s) \equiv f_3(s)+f_9(s)$ with $f_{1,2,3,6,7,9}(s)$ on the right hand sides given in eqs.(2-12) of ref.\cite{subsubvmed}.  
\subsection{Contributions to in-medium NN-potential}
Again, one lists the four contributions from $V_\text{3n}$ in eq.(8) to the in-medium nn-potential $V_\text{med}$. The self-closing of the neutron lines 1 and 2 lead to nonvanishing contributions, and with the relations $f_3(0) =5(1-g_A^2)m_\pi $ and $f_5(q)+q^2f_4(q)=0$ one obtains: 
\begin{equation}V^{(0)}_\text{med} = {5g^4_A (1-g_A^2)m_\pi k^3_n \over 384\pi^3 f_\pi^6}\,
{\vec\sigma_1\!\cdot\!\vec q\, \vec\sigma_2\!\cdot\!\vec q \over m^2_\pi+q^2}\,, \end{equation}
which is of the form: $1\pi^0$-exchange nn-interaction times a factor linear in the neutron density $\rho_n=k_n^3 /3\pi^2$. On the other hand the vertex corrections by $1\pi^0$-exchange, incorporated in eq.(8) through the second factor $\vec\sigma_3\!\cdot\! \vec q_3/(m_\pi^2+q_3^2)$, produce the contribution:
\begin{eqnarray}V^{(1)}_\text{med} &=&  {g_A^4 \over (8\pi f_\pi^2)^3} \bigg\{ 
\Big(2m_\pi^2\Gamma_0-{4k_n^3\over 3}\Big) f_3(q) -\Big(2\Gamma_2 +{q^2 \over 2} \widetilde\Gamma_3\Big)  q^2 f_1(q)+ \widetilde\Gamma_1  q^2 f_2(q)-2\vec\sigma_1\!\cdot\!\vec\sigma_2\,\Gamma_2 f_5(q)\nonumber\\ && 
- (\vec\sigma_1\!\cdot\!\vec p\, \vec\sigma_2\!\cdot\!\vec p+\vec\sigma_1\!\cdot\!\vec p\,'\, \vec\sigma_2\!\cdot\!\vec p\,')\widetilde \Gamma_3 f_5(q) -\vec\sigma_1\!\cdot\!\vec q\, \vec\sigma_2\!\cdot\!\vec q\,\Big(2\Gamma_2 +{q^2 \over 2} \widetilde\Gamma_3\Big) f_4(q) \nonumber\\ && +i( \vec\sigma_1\!+\!\vec\sigma_2)\!\cdot \! (\vec q\!\times \! \vec p\,) \Big[{q^2 \over 2}\widetilde\Gamma_3 f_1(q)- \widetilde\Gamma_1 f_2(q) \Big] +\Big({2k_n^3 \over 3}-m_\pi^2 \widetilde \Gamma_1\Big) q^2 f_8(q)\bigg\}\,. \end{eqnarray}
Here, the frequently occuring combinations $\widetilde\Gamma_1(p)= \Gamma_0(p)+\Gamma_1(p)$ and $\widetilde\Gamma_3(p)= \Gamma_0(p) +2\Gamma_1(p)+\Gamma_3(p)$ have been introduced.  Moreover, the vertex corrections by $2\pi$-exchange (compiled in the expression in curly brackets of eq.(8)) can be summarized as the $1\pi^0$-exchange nn-interaction times a $(p,q,k_f)$-dependent factor: 
\begin{equation}V^{(2)}_\text{med} = {g^4_A \over (8\pi f^2_\pi)^3} \,{\vec\sigma_1\!\cdot\!\vec
q\, \vec\sigma_2\!\cdot\!\vec q \over m^2_\pi+q^2}\,\big[ S_1(p) +q^2 S_2(p)\big]\,, \end{equation}
with the two functions $S_{1,2}(p,k_n)$ defined by:
\begin{eqnarray}S_1(p)&=& {1\over 8 p^3}\int_{p-k_n}^{p+k_n}\!\!\!ds\, s\big[k_n^2-(p-s)^2\big] \Big\{\big[(p+s)^2-k_n^2)\big]f_2(s)-8p^2\big[f_3(s)+f_5(s)\big]\nonumber\\ && +{1\over 3}\big[k_n^2-(p-s)^2\big] (k_n^2-p^2-4p s -s^2) \big[f_1(s)+f_4(s)\big]\Big\}\,,  \end{eqnarray}
\begin{equation}S_2(p)= {1\over 32 p^5}\int_{p-k_n}^{p+k_n}\!\!\!ds\, s\big[k_n^2-(p-s)^2\big]\big[k_n^2-(p+s)^2\big]\Big\{(p^2+s^2-k_n^2)\big[f_1(s)+f_4(s)\big]-4p^2 f_8(s)\Big\}\,. \end{equation}
Note that the integrals for $S_{1,2}(p)$ with input functions $f_{1,2,3,4,5}(s)$ as in eqs.(9-12) could be solved in terms of the functions $\arctan[(p\pm k_n)/ 2m_\pi]$ and $ \ln[4m_\pi^2+(p\pm k_n)^2]$.

Finally, the more complicated contribution from double exchange  reads:
\begin{eqnarray}V^{(3)}_\text{med}&\!\!=\!\!&{g_A^4\over (8\pi f_\pi^2)^3}\bigg\{ 
\vec\sigma_1\!\cdot\!\vec \sigma_2\big(2I_{2,2}-2I_{3,2}-H_{1,2}-\tilde I_{1,2}+H_{8,2}+\tilde I_{8,2}\big)+ \vec\sigma_1\!\cdot\!\vec q\, \vec \sigma_2\!\cdot\!\vec q \,\Big({H_{1,1}+\tilde I_{1,4}+\tilde I_{8,5}\over 2}\nonumber\\ && -I_{2,4}-I_{3,5}+H_{8,0}-H_{8,1}\Big)+(\vec\sigma_1\!\cdot\!\vec p\, \vec\sigma_2\!\cdot\!\vec p+\vec\sigma_1\!\cdot\!\vec p\,'\, \vec\sigma_2\!\cdot\!\vec p\,')\Big(I_{2,3}-I_{3,3}-{H_{1,3}+\tilde I_{1,3}\over 2}\nonumber\\ && +{H_{8,3}+\tilde I_{8,3}\over 2}\Big) + 2m_\pi^2 I_{5,0}-2H_{5,0}-3H_{4,2}-3 \tilde I_{4,2} +{q^2 \over 2}(H_{4,1}+\tilde I_{4,4})-p^2(H_{4,3}+\tilde I_{4,3})\nonumber\\ && -{i\over 2}(\vec\sigma_1 \!+\!\vec\sigma_2)\!\cdot\!(\vec q\!\times\!\vec p\,)(H_{4,1}+\tilde I_{4,1})\bigg\}\,.  \end{eqnarray}
The definitions of the double-indexed functions $H_{j,\nu}(p,k_n),\, I_{j,\nu}(p,q,k_n)$ and $\tilde I_{j,\nu}(p,q,k_n)$ can be found in eqs.(25-34) of ref.\cite{vmedlong}.

 One should point to a typing error in the third line of eq.(24) in ref.\cite{vmedlong}. There the spin-orbit operator should read  $i(\vec\sigma_1 \!+\!\vec\sigma_2)\!\cdot\!(\vec q\!\times\!\vec p\,)$ and not $i(\vec\sigma_1 \!+\!\vec\sigma_2)\!\cdot\!(\vec p\!\times\!\vec q\,)$. The same typing error has occurred in the second line of eq.(19) in ref.\cite{subsubvmed}. Further typing errors that need to be corrected in ref.\cite{vmedshort} are: $(q^2/4-2p^2)$ at the beginning of the last line of eq.(29), $[m_\pi^2(\gamma_0+\gamma_1)-\Gamma_0-\Gamma_1]$ at the end of the fourth line of eq.(30), and $(\dots +G_{0*}+2
G_{1*})$ in the fifth line of eq.(45). 

\section{Ring topology}
The three-neutron ring-interaction, represented by the right diagram in Fig.\,1,  possesses a more complicated structure, because any factorization property in the three momentum transfers $\vec q_{1,2,3}$ is lost. One starts with the basic expression for $ V_\text{3n}$ in the form of a three-dimensional loop-integral over (static) pion-propagators and momentum-factors \cite{3Nlong}:
\begin{eqnarray}V_\text{3n}&\!\!\!\!=\!\!\!\!&{g_A^4\over 16f_\pi^6}\int\!{d^3l_2\over (2\pi)^3} {1\over (m_\pi^2+l_1^2)(m_\pi^2+l_2^2)(m_\pi^2+l_3^2)} \bigg\{\vec l_1\!\cdot\!\vec l_2\, \vec l_2\!\cdot\!\vec l_3 -\vec\sigma_1\!\cdot\!(\vec l_2\!\times\!\vec l_3) \vec\sigma_3\!\cdot\!(\vec l_1\!\times\!\vec l_2)\nonumber\\ &&-{g_A^2 \over m_\pi^2+l_2^2} \Big[ \vec l_1\!\cdot\!\vec l_2\, \vec l_1\!\cdot\!\vec l_3\,\vec l_2\!\cdot\!\vec l_3 + 2\vec\sigma_2\!\cdot\!(\vec l_1\!\times\!\vec l_3) \vec\sigma_3\!\cdot\!(\vec l_1\!\times\!\vec l_2)\vec l_2\!\cdot\!\vec l_3  -{3\over 2} \vec\sigma_1\!\cdot\!(\vec l_2\!\times\!\vec l_3)\vec\sigma_3\!\cdot\!(\vec l_1\!\times\!\vec l_2)\vec l_1\!\cdot\!\vec l_3\Big]\bigg\} \,, \end{eqnarray}
where one has to set $\vec l_1= \vec l_2-\vec q_3$ and $\vec l_3= \vec l_2+\vec q_1$. 

\subsection{Self-closings of neutron-lines} 
For the self-closings of neutron-lines the Fermi-sphere integral gives just a factor density $\rho_n= k_n^3/3\pi^2$. Sorted according to powers of $g_A^2$, the self-closing contributions to $V_\text{med}$ from the 3n-ring interaction $V_\text{3n}$ in eq.(19) read: 
\begin{eqnarray}
V^{(0)}_\text{med} &=& {g^4_A k_n^3 \over 192\pi^3f_\pi^6}\bigg\{-7m_\pi-{m_\pi^3\over 4m_\pi^2+q^2}-{8m_\pi^2+3q^2\over q}\arctan{q\over 2m_\pi}\nonumber \\ && +\big(\vec\sigma_1\!\cdot\!\vec\sigma_2\,q^2-\vec\sigma_1\!\cdot\!\vec q\, \vec\sigma_2\!\cdot\!\vec q\,\big){1\over 2q} \arctan{q\over 2m_\pi}\bigg\}\,, \end{eqnarray}
\begin{eqnarray}
V^{(0)}_\text{med}& = &{g^6_A k_n^3\over 192\pi^3f_\pi^6}\bigg\{{19m_\pi\over 2}-{5m_\pi^3 \over 2(4m_\pi^2+q^2)}+{4m_\pi^5 \over (4m_\pi^2+q^2)^2}+{8m_\pi^2+3q^2\over q}\arctan{q\over 2m_\pi} \nonumber\\ &&+ {1\over 2q^2}\big(\vec\sigma_1\!\cdot\!\vec \sigma_2\,q^2 -\vec\sigma_1\!\cdot\!\vec q\,\vec\sigma_2\!\cdot\!\vec q\, \big)\bigg[m_\pi-{5m_\pi^3\over 4m^2_\pi+q^2}+{m_\pi^2-7q^2\over 2q}\arctan{q \over 2m_\pi}\bigg]\bigg\}\,,  \end{eqnarray} 
where the linear divergences in the central parts have been treated by dimensional regularization. 
\subsection{Concatenations of neutron-lines for ring interaction $\sim g_A^4$}
Treating the concatenations of two neutron-lines is somewhat simpler for the $g_A^4$-part of the 3n-ring interaction $V_\text{3n}$ in eq.(19), since for this component the three pion-propagators are on an equal footing. Using the assignments of momenta $\vec l_1,\vec l_2,\vec l_3$ for the six possible concatenations in table\,1 of ref.\cite{vmedlong}, one obtains the following contributions to $V_\text{med}$, which are listed individually by specifying first their type. 
\smallskip
\noindent $\bullet$ Central potential (three pieces added):
\begin{eqnarray} V^\text{(cc)}_\text{med} &\!=\!& {g_A^4 \over 64\pi^4 f_\pi^6} \int_0^\infty\!\!\!dl\bigg\{ l \widetilde\Gamma_1(l)\bigg[l\Big({q^2\over 2p^2}-2\Big) +\Big({q^2\over 2}+4m_\pi^2-2l^2+2p^2-{q^2\over 2p^2}(m_\pi^2+l^2) \Big) \Lambda(l)  \nonumber \\ &&+ (l^2-m_\pi^2-p^2)(2m_\pi^2+q^2) \Omega(l)\bigg]+l\bigg[{2k_n^3\over 3}-m_\pi^2\Gamma_0(l)\bigg]\Big[(2m_\pi^2+q^2) \Omega(l)-2\Lambda(l)\Big]+4k_n^3 \bigg\}\,, \nonumber \\ &&\end{eqnarray}
with the following auxiliary functions arising from the solid-angle integration:
\begin{equation} \Lambda(l)={1\over 4p}\ln {m_\pi^2+(l+p)^2 \over m_\pi^2+(l-p)^2}\,,  \end{equation}
\begin{equation} \Omega(l)={1\over q \sqrt{B+q^2l^2}}\ln {q\, l +\sqrt{B+q^2l^2} \over \sqrt{B}}\,,  \end{equation}
and the abbreviation $B= [m_\pi^2+(l+p)^2][m_\pi^2+(l-p)^2]$. In accordance with dimensional regularization the subtraction of the asymptotic constant $-4k_n^3$ ensures the convergence of the radial integral $\int_0^\infty\!dl$ in eq.(22).

\noindent $\bullet$ Spin-spin and tensor potentials (two pieces added):
\begin{equation} V^\text{(cc)}_\text{med}={g_A^4\over 32\pi^4 f_\pi^6}(\vec\sigma_1\!\cdot\!\vec \sigma_2 q^2 -\vec\sigma_1\!\cdot\!\vec q\,\vec\sigma_2\!\cdot\!\vec q\,) \!\int_0^\infty\!\!\!dl{l \widetilde\Gamma_1(l)\over 4p^2-q^2} \Big[(B+q^2l^2) \Omega(l)-(m_\pi^2+l^2+p^2)\Lambda(l)\Big]\,, \end{equation}
\noindent $\bullet$ Quadratic spin-orbit potential (two pieces added):
\begin{eqnarray}
V^\text{(cc)}_\text{med}&=&{g_A^4\over 16\pi^4 f_\pi^6} \,\vec\sigma_1\!\cdot\!(\vec q \!\times \! \vec p\,)\vec\sigma_2\!\cdot\!(\vec q \!\times \! \vec p\,) \!\int_0^\infty\!\!\!dl{l \widetilde\Gamma_1(l)\over 4p^2-q^2} \bigg\{-{l \over 2p^2}+ \bigg[{m_\pi^2+l^2-p^2\over 2p^2}\nonumber \\ && +{4(m_\pi^2+l^2+p^2)\over 4p^2 -q^2}\bigg]\Lambda(l) +\bigg[m_\pi^2+3l^2+p^2 -{4( m_\pi^2+l^2+p^2)^2 \over 4p^2-q^2}\bigg] \Omega(l)\bigg\} \,, \end{eqnarray}
\noindent $\bullet$ Spin-orbit term (one piece):
\begin{eqnarray}
V^\text{(cc)}_\text{med}&=&{g_A^4\over 64\pi^4 f_\pi^6}\, i(\vec\sigma_1\!+\!\vec\sigma_2)\!\cdot\!(\vec q \!\times \! \vec p\,)\!\int_0^\infty\!\!\!dl{l \over 4p^2-q^2}\bigg\{ \Gamma_2(l)\Big[(2m_\pi^2+2l^2-2p^2+q^2)\Omega(l)- 2\Lambda(l)\Big]\nonumber \\ && \qquad\qquad\qquad\qquad\qquad\qquad\quad  +\widetilde\Gamma_3(l) \Big[ (B+q^2l^2)\Omega(l)-(m_\pi^2+l^2+p^2) \Lambda(l)\Big] \bigg\}\,.  \end{eqnarray}

\subsection{Concatenations of neutron-lines for ring interaction $\sim g_A^6$}
When treating the concatenations of two neutron-lines for the $g_A^6$-part of the 3n-ring interaction $V_\text{3n}$, one has to distinguish the cases with $\vec l_2 =\pm(\vec l_4+\vec l\,)$, because this momentum enters a squared pion-propagator in eq.(19). The Fermi-sphere integral over the squared pion-propagator (yielding the functions $\gamma_\nu(l,k_n)$) combined with the solution of the angular integral leads to the following contributions to $V_\text{med}$ from the concatenations $n_1$ on $n_3$ and $n_3$ on $n_1$.

\noindent $\bullet$ Central potential:
\begin{eqnarray} V^\text{(2)}_\text{med} &=& {g_A^6 \over 64\pi^4 f_\pi^6} \int_0^\infty\!\!\!dl\bigg\{ 2l \gamma_2(l)\bigg[l\Big(4-{q^2\over p^2}\Big)+\Big((m_\pi^2\!+\!l^2) {q^2\over p^2} -8m_\pi^2-3q^2\Big)\Lambda(l)+ (2m_\pi^2\!+\!q^2)^2\, \Omega(l)\bigg]\nonumber \\ && + {l\over 8} \widetilde \gamma_3(l)\bigg[
2l(2m_\pi^2+9l^2+p^2)-{l q^2\over p^2}(m_\pi^2+5l^2) +(2m_\pi^2+q^2) \big[B+2l^2(4m_\pi^2+3q^2)\big] \Omega(l)\nonumber \\ && +\Big[{q^2 \over p^2}(m_\pi^2+l^2)(m_\pi^2+5l^2) -q^2(16l^2+p^2)-2B-4m_\pi^2(m_\pi^2+9l^2+p^2) \Big] \Lambda(l)\bigg] -{4k_n^3 \over 3}\bigg\}\,,\nonumber \\ &&  \end{eqnarray}
\noindent $\bullet$ Spin-orbit potential:
\begin{eqnarray} V^\text{(2)}_\text{med} &=& {3g_A^6 \over (4\pi)^4 f_\pi^6} i (\vec\sigma_1+\vec\sigma_2)\!\cdot\!(\vec q \!\times \! \vec p\,)\!\int_0^\infty\!\!\!dl{l \over 4p^2-q^2}\bigg\{\gamma_2(l)\bigg[\Big[ 4(p^2\!-\!l^2\!-\!2m_\pi^2)-3q^2 +{q^2\over p^2}(m_\pi^2\!+\!l^2)\Big]\Lambda(l) \nonumber \\ && +l \Big(4-{q^2\over p^2}\Big) +(2m_\pi^2\!+\!q^2)(2m_\pi^2+2l^2-2p^2+q^2)\Omega(l)\bigg] +\widetilde\gamma_3(l)\bigg[ l\Big(1-{q^2 \over 4p^2}\Big)(m_\pi^2\!+\!l^2\!+p^2) \nonumber \\ && +\Big[{q^2 \over 4p^2}(m_\pi^2\!+\!l^2)^2 -B-2m_\pi^2(m_\pi^2+l^2+p^2)-{q^2\over 4}(2m_\pi^2+6l^2+3p^2) \Big] \Lambda(l) \nonumber \\ && +( 2m_\pi^2+q^2)(B+q^2l^2) \Omega(l)\bigg] \bigg\}\,, \end{eqnarray}
\noindent $\bullet$ Spin-spin and tensor potentials:
\begin{eqnarray} V^\text{(2)}_\text{med} &=& {g_A^6 \over  16\pi^4 f_\pi^6}(\vec\sigma_1\!\cdot\!\vec \sigma_2 q^2 -\vec\sigma_1\!\cdot\!\vec q\,\vec\sigma_2\!\cdot\!\vec q\,) \!\int_0^\infty\!\!\!dl{l \over 4p^2-q^2}\bigg\{ \gamma_2(l) \Big[(m_\pi^2+l^2+p^2) \Lambda(l)-(B+q^2l^2)  \Omega(l)\Big] \nonumber \\ && + {\widetilde\gamma_3(l)\over 8} \bigg[l \Big(1-{q^2\over 4p^2}\Big)(m_\pi^2+l^2+p^2)+\Big[(l^2+p^2-m_\pi^2)^2+{q^2 \over 4}(2m_\pi^2-2l^2+p^2) \nonumber \\ && -4m_\pi^4-4m_\pi^2p^2-4p^4+{q^2 \over 4p^2}(m_\pi^2+l^2)^2\Big] \Lambda(l) +2(m_\pi^2-l^2+p^2) (B+q^2l^2)  \Omega(l)\bigg]\bigg\}\,, \end{eqnarray}
\noindent $\bullet$ Quadratic spin-orbit potential:
\begin{eqnarray}
V^\text{(2)}_\text{med}&=&{g_A^6 \over 16\pi^4 f_\pi^6} \,\vec\sigma_1\!\cdot\!(\vec q \!\times \! \vec p\,)\vec\sigma_2\!\cdot\!(\vec q \!\times \! \vec p\,) \!\int_0^\infty\!\!\!dl{l\over 4p^2-q^2} \bigg\{ \gamma_2(l)\bigg[ {l \over p^2}+\Big[3-{ m_\pi^2+l^2\over p^2} \nonumber \\ &&-{8(m_\pi^2+l^2+p^2)\over 4p^2-q^2} \Big] \Lambda(l) +\Big[ {8(m_\pi^2+l^2+p^2)^2 \over 4p^2-q^2}-4m_\pi^2-8l^2-q^2\Big] \Omega(l) \bigg]\nonumber \\ &&  + \widetilde\gamma_3(l)(m_\pi^2+p^2-l^2)\bigg[-{l \over 4p^2}+\Big[ 
{m_\pi^2+l^2-p^2 \over 4p^2}+{2(m_\pi^2+l^2+ p^2) \over 4p^2-q^2} \Big] \Lambda(l) \nonumber \\ && +\Big[{1\over 2}(m_\pi^2+3l^2+p^2)-{2(m_\pi^2+l^2+ p^2)^2 \over 4p^2-q^2}\Big] \Omega(l)\bigg]\bigg\}\,. \end{eqnarray}
For the other four concatenations one has $\vec l_{1,3} = \pm(\vec l_4+\vec l\,)$ and the Fermi-sphere integral goes over an ordinary pion-propagator (yielding the functions $\gamma_\nu(l,k_n)$). When adding pieces of the same type,  one obtains the following additional contributions to $V_\text{med}$ from concatenations. 

\noindent $\bullet$ Central potential (two pieces added)
\begin{eqnarray} V^\text{(cc)}_\text{med}&=&{g_A^6\over 128\pi^4 f_\pi^6} \!\int_0^\infty\!\!\!dl \bigg\{ l\Big[{2k_n^3\over 3}-m_\pi^2 \Gamma_0(l)\Big] \bigg[\Big(8-{q^2\over p^2}\Big) \Lambda(l) \nonumber \\ && +
 {l\over B}\Big(  {q^2 \over p^2}(m_\pi^2 +l^2) - 8m_\pi^2-3q^2  +{(2m_\pi^2 +q^2)^2\over B+q^2l^2} (m_\pi^2+l^2+p^2) \Big) \nonumber \\ &&+(2m_\pi^2 +q^2)\bigg({2m_\pi^2 +q^2\over B+q^2l^2} (m_\pi^2+l^2+p^2) -4\Big)\Omega(l) \bigg]-{16k_n^3\over 3}\bigg\}\,,  \end{eqnarray}
\noindent $\bullet$  Spin-orbit potential (one piece):
\begin{eqnarray} V^\text{(cc)}_\text{med} &=& {g_A^6 \over  128\pi^4 f_\pi^6} i (\vec\sigma_1\!+\!\vec\sigma_2)\!\cdot\!(\vec q \!\times \! \vec p\,)\!\int_0^\infty\!\!\!dl\,l\bigg\{\Gamma_2(l)\bigg[\Big({1 \over p^2}+{4 \over 4p^2-q^2}\Big) \Lambda(l)\nonumber \\ &&+ {l \over B}\Big( 1-{m_\pi^2+l^2\over p^2}+{2m_\pi^2+q^2\over B+q^2l^2} (l^2-m_\pi^2-p^2)\Big) \nonumber \\ &&+\Big[2-{4(m_\pi^2 +l^2+p^2)\over 4p^2-q^2} +{2m_\pi^2+q^2\over B+q^2l^2}(l^2-m_\pi^2-p^2)\Big] \Omega(l) \bigg] \nonumber \\ && + \widetilde\Gamma_3(l)\bigg[ -{l\over 2p^2} +\Big( {m_\pi^2+l^2-p^2\over 2p^2} +{2(2m_\pi^2+l^2+3p^2)\over 4p^2-q^2}\Big)\Lambda(l)
\nonumber \\ && - \Big(2B+(m_\pi^2+p^2)(2m_\pi^2+q^2) +l^2(2m_\pi^2+3q^2) \Big) {\Omega(l)\over 4p^2-q^2} \bigg] \bigg\}\,. \end{eqnarray}
\noindent $\bullet$ Spin-spin and tensor potentials (two pieces added):
\begin{eqnarray} V^\text{(cc)}_\text{med} &=& {g_A^6 \over 64\pi^4 f_\pi^6}(\vec\sigma_1\!\cdot\!\vec \sigma_2 q^2 -\vec\sigma_1\!\cdot\!\vec q\,\vec\sigma_2\!\cdot\!\vec q\,) \!\int_0^\infty\!\!\!dl{l \over 4p^2-q^2}\bigg\{ 5\Gamma_2(l)\Big[(m_\pi^2+l^2+ p^2)\Omega(l)-\Lambda(l)\Big] \nonumber \\ && +{\widetilde\Gamma_3(l)\over 4}\bigg[3l\Big({q^2 \over 2p^2}-2\Big)+\Big[5(3m_\pi^2+l^2+ 3p^2) -{3q^2 \over 2p^2}(m_\pi^2+l^2+p^2)\Big] \Lambda(l) \nonumber \\ && + \Big[ 4l^2(2p^2-2m_\pi^2-q^2)+l^4 -9(m_\pi^2+p^2)^2\Big] \Omega(l)\bigg] \bigg\}\,, \end{eqnarray}
\noindent $\bullet$ Quadratic spin-orbit potential (two pieces added):
\begin{eqnarray}
V^\text{(cc)}_\text{med}&=&{g_A^6 \over 64\pi^4 f_\pi^6} \,\vec\sigma_1\!\cdot\!(\vec q \!\times \! \vec p\,)\vec\sigma_2\!\cdot\!(\vec q \!\times \! \vec p\,) \!\int_0^\infty\!\!\!dl{l\over 4p^2-q^2} \bigg\{ 5\Gamma_2(l)\bigg[\Big( {1\over p^2}+{8 \over 4p^2-q^2}\Big)\Lambda(l) -{l\over B}
\Big[{m_\pi^2 +l^2\over p^2}\nonumber \\ && +{4m_\pi^2+q^2 \over B+q^2l^2}( m_\pi^2 +l^2+ p^2)-3\Big] +\Big[4- ( m_\pi^2 +l^2+ p^2)\Big( {8 \over 4p^2-q^2}+{4m_\pi^2+q^2 \over B+q^2l^2}\Big)\Big]
 \Omega(l) \bigg] \nonumber \\ && +  \widetilde\Gamma_3(l)\bigg[ 2\Big({l^2-9m_\pi^2-9p^2 \over 4p^2-q^2}-
 { m_\pi^2\over p^2} \Big) \Lambda(l) +{l\over B}\Big[{m_\pi^4-(l^2-p^2)^2 \over q^2}
 +{2m_\pi^2\over p^2}(m_\pi^2+l^2) \nonumber \\ && -{11m_\pi^2 \over 2}+{5\over 2}(l^2+p^2-q^2) +{5(4m_\pi^2+q^2) \over 2( B+q^2l^2)} \big((m_\pi^2+p^2)^2+l^2(m_\pi^2-3p^2+q^2)\big) \Big]  \nonumber \\ &&+\Big[
 {(l^2-p^2)^2 -m_\pi^4\over q^2} +{2(m_\pi^2+l^2+p^2)\over 4p^2-q^2}(9m_\pi^2+9 p^2-l^2)-{l^2 \over 2}
  \nonumber \\ &&-{5\over 2}(5m_\pi^2+p^2+q^2) +{5(4m_\pi^2+q^2) \over 2(B+q^2l^2)}\big((m_\pi^2+p^2)^2+l^2(m_\pi^2-3p^2+q^2)\big) \Big] \Omega(l)\bigg] \bigg\}\,.\end{eqnarray}
\section{$2\pi^0$-exchange three-neutron force at N$^4$LO}
In this section the longest-range $2\pi^0$-exchange 3n-interaction is treated, following the work of ref.\cite{twopi4}. Modulo terms of shorter range it can be written according to eq.(3.1) in ref.\cite{twopi4} in the general form:
\begin{equation}2V_\text{3n}={g_A^2 \over 4f_\pi^4} \,{\vec\sigma_1\!\cdot\! \vec
q_1 \vec \sigma_3\!\cdot\! \vec q_3 \over (m_\pi^2+q_1^2) (m_\pi^2+q_3^2)}
\,\tilde g_+(q_2) \,, \end{equation}
with both sides multiplied by a factor $2$ due to the $1\!\leftrightarrow\!3$ symmetry. The structure function $\tilde g_+(q_2)$ is $f_\pi^2$ times the isoscalar non-spin-flip $\pi N$-scattering amplitude at zero pion-energy $\omega =0$ and squared momentum-transfer $t= -q_2^2$. The corresponding expression up to N$^4$LO is given in eq.(59) of ref.\cite{vmedlong}. 

\subsection{Contributions to in-medium nn-potential}
The self-closing of neutron-line $2$ for $V_\text{3n}$ in eq.(36) gives (after relabeling $3\to 2$) the contribution:
\begin{equation}V^{(0)}_\text{med} = -{g^2_A k^3_n \over 12\pi^2 f^4_\pi}\,{\vec\sigma_1\!\cdot\!\vec
q\, \vec\sigma_2\!\cdot\!\vec q\over (m^2_\pi+q^2)^2}\, \tilde g_+(0)\,. \end{equation}
 From pionic vertex corrections on either neutron-line one obtains the (total) contribution:
\begin{eqnarray} V^{(1)}_\text{med} & = &{g^2_A\over 16\pi^2 f^4_\pi}\, {\vec\sigma_1\!\cdot\!\vec
q\, \vec\sigma_2\!\cdot\!\vec q \over (m^2_\pi+q^2)q^2} \int_{p-k_n}^{p+k_n}\!\! ds\, s \,\tilde g_+(s)\bigg[ {k_n^2-(p-s)^2\over p}\nonumber \\ &&  +\Big(q-{m_\pi^2+s^2\over q}\Big) \ln{ q X + 2\sqrt{W} \over (2p+q)[m_\pi^2+(q-s)^2]} \bigg] \,,  \end{eqnarray}
with the auxiliary polynomials $X$ and $W$ defined in eq.(6). Finally, the two diagrams related to double exchange lead to the expression: 
\begin{equation} V^{(3)}_\text{med}  = {g^2_A\,\tilde g_+(q)\over 16\pi^2 f^4_\pi} \Big[
 2\Gamma_0-(2m_\pi^2+q^2) G_0+ i( \vec\sigma_1\!+\!\vec\sigma_2)\!\cdot\!(\vec q\!\times \! \vec p\,)(G_0+2G_1) \Big]\,,  \end{equation}
where one reminds that the functions $\Gamma_0$ depends on $(p,k_n)$, and the functions $G_{0,1}$ depend on $(p,q,k_n)$. 

\section{Subleading ring diagrams}
\subsection{3n-ring interaction $\sim g_A^0$}
According to eq.(31) in ref.\cite{subsubvmed} the subleading 3n-ring interaction proportional to $g_A^0$ reads:
\begin{eqnarray}V_\text{3n}&=&-{1\over f_\pi^6}\int_0^\infty\!\! dl_0\!\int\!{ d^3l_2\over (2\pi)^4} {l_0^2\over (\bar m^2+l_1^2)(\bar m^2+l_2^2)(\bar m^2+l_3^2)} \Big[2c_1 m_\pi^2+(c_2+c_3)l_0^2 +c_3 \,\vec l_2\!\cdot\!\vec l_3\Big]\,, \end{eqnarray}
with $\bar m = \sqrt{m_\pi^2+l_0^2}$ and one has to set $\vec l_1= \vec l_2-\vec q_3$ and $\vec l_3= \vec l_2+\vec q_1$. The sum of the three self-closings of a neutron-line gives rise to the central potential:
\begin{eqnarray} V^{(0)}_\text{med}& = & {k_n^3 \over 32\pi^4 f_\pi^6} \bigg\{ \bigg[ \Big(2c_1-{3c_2\over 2}-3c_3\Big) m_\pi^2 - \Big({c_2\over 4}+{5c_3 \over 9}\Big) q^2\bigg] \ln{m_\pi \over \lambda}  +\Big({5c_2\over 8}+{37c_3 \over 27}\Big){q^2 \over 4}\nonumber \\ && +\Big({3c_2\over 2}-2c_1+{23c_3\over 9}\Big) {m_\pi^2\over 4} +\bigg[\Big(2c_1-c_2-{17c_3\over 9}\Big) m_\pi^2-\Big({c_2\over 4}+{5c_3 \over 9}\Big) q^2\bigg] L(q) \bigg\}\,, 
\end{eqnarray} 
with logarithmic loop-function \begin{equation}L(q) = {\sqrt{4m_\pi^2+q^2}\over q} \ln{q+\sqrt{4m_\pi^2+q^2}\over 2m_\pi}\,.  \end{equation} Note that the four-dimensional euclidean loop-integral in eq.(40) has been regularized with a cutoff $\lambda$ and the $\lambda^2$-divergence has been dropped in the expression in eq.(41). The total contribution from the six possible concatenations of two neutron-lines is a central potential with the following double-integral represenatation:
\begin{eqnarray} V^{(\text{cc})}_\text{med}& \!\!=\!\! & {1 \over 4\pi^5 f_\pi^6}\int_0^\lambda\!\!dr r\!\int_0^{\pi/2}\!\!d\psi \bigg\{ l_0^2l\Big\{ \bar\Gamma_0(l) \Big[ c_3 \bar \Lambda(l) +\Big(6c_1 m_\pi^2+ 3(c_2+c_3)l_0^2-{c_3 \over 2}(2\bar m^2+q^2)\Big)\bar\Omega(l)\Big]  \nonumber \\ && +c_3 \tilde{\bar \Gamma}_1(l) \Big[\bar \Lambda(l)+(l^2-\bar m^2-p^2) \bar \Omega(l)\Big] \Big\} -{k_n^3 \over 2}(c_3+c_2\cos^2\psi)\sin^2 2\psi\bigg\}\,, \end{eqnarray} 
where one has to set $l_0 = r\cos\psi$ and $l = r\sin\psi$. Note that all barred functions are to be evaluated with $\bar m = \sqrt{m_\pi^2+l_0^2}$ instead of $m_\pi$. The behavior of the subtracted double-integral for large $\lambda$ is:
\begin{equation} {\pi k_n^3 \over 16}\bigg[ c_2\Big( 3m_\pi^2+{3k_n^2 \over 10}+{p^2\over 2}+{q^2 \over 4}\Big) -4c_1 m_\pi^2+c_3\Big( 6m_\pi^2+{2k_n^2 \over 3}+{5\over 9}(2p^2+q^2)\Big)\bigg] \ln {m_\pi\over \lambda} \,.\end{equation} 
A good check is given by the fact that the $\lambda^2$-divergences behind $V^{(0)}_\text{med}$ and $V^{(\text{cc})}_\text{med}$ cancel each other. This feature must hold, because there is no 3n-contact coupling that could absorb this divergence.
\subsection{3n-ring interaction  $\sim g_A^2$}
According to eq.(44) in ref.\cite{subsubvmed} the subleading 3n-ring interaction proportional to $g_A^2$ reads:
\begin{eqnarray}V_\text{3n}&\!\!=\!\!&-{g_A^2\over f_\pi^6}\int_0^\infty\!\! dl_0\!\int\!{ d^3l_2\over (2\pi)^4} {1\over (\bar m^2+l_1^2)(\bar m^2+l_2^2)(\bar m^2+l_3^2)} \Big\{ \big[2c_1 m_\pi^2+(c_2+c_3)l_0^2 +c_3 \,\vec l_2\!\cdot\!\vec l_3\big]\, \vec l_1\!\cdot\!(\vec l_2+\vec l_3) \nonumber\\ && +\,c_4\Big[ \bar m^2(\vec \sigma_2\!\times\!\vec l_3)\!\cdot\!(\vec \sigma_3\!\times\!\vec l_2)  +\vec l_2\!\cdot\!\vec l_3\, \vec \sigma_2\!\cdot\!\vec l_1\, \vec \sigma_3\!\cdot\!\vec l_1 + \vec l_1\!\cdot\!\vec l_2\,\vec l_1\!\cdot\!\vec l_3 \, \vec \sigma_2\!\cdot\!\vec \sigma_3-2 \vec l_1\!\cdot\!\vec l_3\, \vec \sigma_2\!\cdot\!\vec l_2\, \vec \sigma_3\!\cdot\!\vec l_1  \Big]  \Big\}\,. \end{eqnarray}
The sum of the three self-closings of a neutron-line gives rise to the following contributions to  $V_\text{med}$.

\noindent $\bullet$ Central potential:
\begin{eqnarray} V^{(0)}_\text{med}& = & {g_A^2k_n^3 \over 96\pi^4 f_\pi^6} \bigg\{ \bigg[ 9m_\pi^2(4c_1-c_2-6c_3)- \Big({11c_2\over 6}+{32c_3 \over 3}\Big) q^2\bigg] \ln{m_\pi \over \lambda} \nonumber \\ &&  +\Big({19c_2\over 12}-5c_1+{25c_3\over 6}\Big) m_\pi^2 +\Big({137c_2\over 16}+58c_3\Big){q^2 \over 9}\nonumber \\ && +\bigg[2m_\pi^2\Big(18c_1-{8c_2\over 3}-{43c_3\over 3}\Big)-\Big({11c_2\over 6}+{32c_3 \over 3}\Big) q^2+{8(c_3-2c_1)m_\pi^4\over 4m_\pi^2+q^2}\bigg] L(q) \bigg\}\,, 
\end{eqnarray} 
\noindent $\bullet$ Spin-spin potential:
\begin{equation} V^{(0)}_\text{med}= {g_A^2c_4k_n^3 \over 96\pi^4 f_\pi^6}\vec\sigma_1\!\cdot\!\vec\sigma_2 \Big\{ -(4m_\pi^2+q^2)\ln{m_\pi \over \lambda} +{q^2\over 4} -m_\pi^2-q^2 L(q)  \Big\}\,, \end{equation}
\noindent $\bullet$ Tensor potential:
\begin{equation} V^{(0)}_\text{med}= {g_A^2c_4k_n^3 \over 96\pi^4 f_\pi^6}\vec\sigma_1\!\cdot\!\vec q\,\vec\sigma_2\!\cdot\!\vec q \,\Big\{ \ln{m_\pi \over \lambda} -{1\over 4} + L(q)  \Big\}\,. \end{equation}
Next, one evaluates the concatenations $n_3$ on $n_2$ and $n_2$ on $n_3$ and obtains the following contributions to $V_\text{med}$.

\noindent $\bullet$ Central potential:
\begin{eqnarray} V_\text{med}^{(1)}&=& {g_A^2\over 4\pi^5 f_\pi^6} \int_0^\lambda\!\! dr r \!\int_0^{\pi/2}\!\!\! d\psi \bigg\{l \tilde{\bar \Gamma}_1(l)\Big\{ c_3 l \Big(1-{q^2\over 4p^2}\Big) +\Big[ 2c_1 m_\pi^2+c_2 l_0^2 +c_3\Big( r^2-2 \bar m^2-p^2\nonumber\\ && +\,{q^2\over 4p^2}(l^2+\bar m^2-p^2)\Big) \Big] \bar\Lambda(l) + \Big[ 2c_1 m_\pi^2+c_2 l_0^2 -c_3\Big(m_\pi^2+{q^2 \over 2}\Big) \Big] (l^2- \bar m^2-p^2) \bar\Omega(l)\Big\} \nonumber\\ && +\, c_4 l \Big\{\big[\bar m^2 \bar \Gamma_0(l)+2 \bar \Gamma_2(l)\big] \big[2 \bar\Lambda(l)- (2\bar m^2+q^2)\bar\Omega(l)\big] +{1\over 4}\tilde {\bar\Gamma}_3(l)\Big[2(3l^2-\bar m^2-p^2) \bar\Lambda(l)\nonumber\\ && +\,l +\big( \bar B-2l^2(4\bar m^2+q^2)\big)\bar\Omega(l)\Big] \Big\}-{4 k_n^3\over 3}(c_3+c_4+c_2 \cos^2\psi) \sin^4\psi\bigg\}\,, \end{eqnarray}
where the (subtracted) double-integral has the large-$\lambda$ behavior: 
\begin{equation} {\pi k_n^3 \over 12 }\bigg[{c_2\over 4}\Big( 6m_\pi^2+k_n^2 +{5p^2 \over 3}+{q^2 \over 6}\Big) -6c_1 m_\pi^2 +c_3\Big(9m_\pi^2+k_n^2+{5p^2 \over 3}+q^2\Big)+c_4\Big(6 m_\pi^2+{3k_n^2\over 5} + p^2+{7q^2\over 6}\Big)\bigg] \ln{m_\pi\over \lambda} \,. \end{equation}
\noindent $\bullet$ Spin-orbit potential:
\begin{eqnarray} V_\text{med}^{(1)}&=& {c_4 g_A^2 \over 8\pi^5 f_\pi^6} i(\vec \sigma_1\!+\!\vec \sigma_2)\!\cdot\!(\vec q\!\times\!\vec p\,) \int_0^\lambda\!\! dr r \!\int_0^{\pi/2}\!\! \!d\psi \, { l\over 4p^2-q^2} \Big\{\big[\bar m^2 \bar \Gamma_0(l)+2 \bar \Gamma_2(l)\big]\big[2 \bar \Lambda(l) \nonumber\\ &&+ (2p^2-2l^2-2\bar m^2-q^2)\bar \Omega(l)\big] + \tilde{\bar \Gamma}_3(l)(\bar m^2+p^2-l^2) \big[(\bar m^2+l^2+p^2)\Omega(l)-\Lambda(l)\big] \Big\}  \,, \end{eqnarray}
with a large-$\lambda$ behavior of the double-integral: $(\pi k_n^3 /24) \ln(m_\pi/ \lambda)$.\\
The other four concatenations ($n_3$ on $n_1$, $n_1$ on $n_3$, $n_1$ on $n_2$, $n_2$ on $n_1$) lead to the following contributions to $V_\text{med}$.

\noindent $\bullet$ Central potential: 
\begin{eqnarray} V_\text{med}^{(\text{cc})}&=& {g_A^2\over 4\pi^5 f_\pi^6} 
\int_0^\lambda\!\! dr r \!\int_0^{\pi/2}\!\!\! d\psi \bigg\{ l  \big[ 2c_1 m_\pi^2+(c_2+c_3) l_0^2\big] 
\Big\{ \bar \Gamma_0(l)\big[2\bar\Lambda(l)-(2\bar m^2+q^2)\bar\Omega(l)\big] \nonumber\\ &&+\tilde{\bar \Gamma}_1(l)\big[\Lambda(l)+(l^2-p^2-\bar m^2)  \Omega(l)\big]\Big\} +c_3 l \Big\{ \bar \Gamma_2(l)\big[ 2\bar\Lambda(l)-(2
\bar m^2+q^2) \bar\Omega(l)\big] \nonumber\\ &&+\tilde{\bar \Gamma}_3(l)\Big[{l \over 2} +(l^2-p^2-\bar m^2)\bar\Lambda(l)+{1\over 2}(l^2-p^2-\bar m^2)^2 \bar\Omega(l)\Big] +\tilde{\bar \Gamma}_1(l)\Big[l \Big(1-{q^2\over 4p^2}\Big) \nonumber\\ &&+\Big( l^2-p^2-2\bar m^2-{q^2 \over 4} +{q^2 \over 4p^2}(l^2+\bar m^2) \Big) \bar\Lambda(l)+\Big(\bar m^2+{q^2\over 2}\Big)(\bar m^2+p^2-l^2) \bar\Omega(l)\Big] \Big\} \nonumber\\ && -{8k_n^3\over 3}(c_3+c_2\cos^2\psi)\sin^4\psi\bigg\}\,, \end{eqnarray}
with a large-$\lambda$ behavior of the (subtracted) double-integral:
\begin{equation} {\pi k_n^3 \over 4 }\bigg[c_2\Big( m_\pi^2+{k_n^2 \over 10}+{p^2 \over 6}+{5q^2 \over 36}\Big)-4c_1 m_\pi^2 +c_3\Big(6m_\pi^2+ {11k_n^2 \over 15}+{11p^2 +5q^2 \over 9}\Big)  \bigg] \ln {m_\pi \over \lambda} \,. \end{equation} 
\noindent $\bullet$ Spin-spin potential:
\begin{eqnarray} V_\text{med}^{(\text{cc})}&=& {c_4 g_A^2 \over 16\pi^5 f_\pi^6} \vec \sigma_1\!\cdot\!\vec \sigma_2\,\int_0^\lambda\!\! dr r \!\int_0^{\pi/2}\!\! \!d\psi \bigg\{ l\tilde{\bar\Gamma}_1(l) \bigg[{l \over p^2}(3 p^2-l^2-\bar m^2 -q^2) \nonumber\\ && + \Big(2l^2-2\bar m^2 -3p^2 +(\bar m^2 +l^2+q^2) {\bar m^2 +l^2\over p^2} + {q^2(4\bar m^2 +4l^2+q^2) \over 4p^2-q^2} \Big)\bar\Lambda(l) \nonumber\\ && + {2q^2(\bar m^2 +l^2+p^2) \over 4p^2-q^2}(2 p^2-2l^2-2\bar m^2 -q^2)  \bar \Omega(l)\bigg] -{32 k_n^3 \over 9}\sin^4\psi \bigg\}  \,, \end{eqnarray}
with a  large-$\lambda$ behavior of the (subtracted) double-integral: $(2\pi k_n^3 /9)\big[ 6m_\pi^2 +k_n^2+2p^2+3q^2/4\big] \ln(m_\pi/ \lambda)$.

\noindent $\bullet$ Ordinary tensor potential:  
\begin{eqnarray} V_\text{med}^{(\text{cc})}&=& {c_4 g_A^2 \over 4\pi^5 f_\pi^6} \vec \sigma_1\!\cdot\!\vec q\, \vec \sigma_2\!\cdot\!\vec q \,\int_0^\lambda\!\! dr r \!\int_0^{\pi/2}\!\! \!d\psi {l \tilde{\bar\Gamma}_1(l)\over 4p^2-q^2}\bigg\{l\Big(1-{q^2\over 2p^2}\Big) \nonumber\\ && + \Big[{q^2\over 4p^2-q^2}\Big(2\bar m^2 +2l^2+{q^2\over 2}\Big)+{\bar m^2+l^2\over 2p^2}(q^2-4p^2)\Big] \bar \Lambda(l)\nonumber\\ && +\Big[3\bar B +2l^2(5p^2+q^2)+(\bar m^2+p^2)(q^2-2p^2) -{8 p^2\over 4 p^2-q^2}(\bar m^2+l^2+p^2)^2 \Big]  \bar\Omega(l) \bigg\}  \,,  \end{eqnarray}
with a large-$\lambda$ behavior of the double-integral: $-(\pi k_n^3 /24)\ln(m_\pi/ \lambda)$.

\noindent $\bullet$ Tensor-type potential:  
\begin{eqnarray} V_\text{med}^{(\text{cc})}&=& {c_4 g_A^2 \over 4\pi^5 f_\pi^6} (\vec \sigma_1\!\cdot\!\vec p\, \vec \sigma_2\!\cdot\!\vec p+\vec \sigma_1\!\cdot\!\vec p\,'\vec \sigma_2\!\cdot\!\vec p{\,'}) \int_0^\lambda\!\! dr r \!\int_0^{\pi/2}\!\! \!d\psi {l \tilde{\bar\Gamma}_1(l)\over 4p^2-q^2} \bigg\{ {l \over 2p^2}\Big[3\bar m^2+3l^2\ -p^2+{5q^2\over 4}\nonumber\\ && -{3q^2\over 4p^2}(\bar m^2+l^2)\Big]+ \bigg[{p^2\over 2}+l^2-\bar m^2-{5q^2\over 8}- {6(\bar m^2+l^2)^2+ q^2(\bar m^2+3l^2) \over 4p^2} +{3q^2\over 8p^4}(\bar m^2+l^2)^2\nonumber\\ && - {q^2(4\bar m^2+4l^2+q^2)\over 4 p^2-q^2}\bigg]\bar\Lambda(l)  +\Big[{4(\bar m^2+l^2+p^2)^2\over 4 p^2-q^2}-\bar m^2-3l^2- p^2\Big]q^2\,\bar \Omega(l) \bigg\}  \,, \end{eqnarray}
with a large-$\lambda$ behavior of the double-integral: $-(\pi k_n^3 /36)\ln(m_\pi/ \lambda)$. 
The total $\lambda^2$-divergence behind the calculated central and spin-spin potentials adds up to $c_4g_A^2 k_n^3\lambda^2/(96\pi^4f_\pi^6) [ 3 +\vec \sigma_1\!\cdot\!\vec \sigma_2]$. As required by the renormalizability condition this is equivalent to zero, since $\vec \sigma_1\!\cdot\!\vec \sigma_2$ has the eigenvalue $-3$ in the $^1S_0$-state of two neutrons.
\subsection{3n-ring interaction $\sim g_A^4$}
In the case of the subleading 3n-ring interaction terms proportional to $g_A^4$, as obtained by setting $\vec\tau_i\!\cdot\!\vec\tau_j\to 1$ and  $\vec\tau_1\!\cdot\!(\vec\tau_2\!\times\!\vec\tau_3) \to 0$ in the very lengthy expression in eq.(58) of ref.\cite{subsubvmed}, only the results for the self-closing contributions are given here.

\noindent $\bullet$ Central potential:
\begin{eqnarray} V^{(0)}_\text{med}& = & {g_A^4k_n^3 \over 96\pi^4 f_\pi^6} \bigg\{ \bigg[ 45m_\pi^2\Big({11c_3-3c_2\over 4}-2c_1\Big)+ (177c_3-53c_2) {q^2\over 8}\bigg] \ln{m_\pi \over \lambda} -{4m_\pi^4 (2c_1+3c_3)\over 4m_\pi^2+q^2}\nonumber \\ &&  +(705c_3-73c_2-528c_1) {m_\pi^2\over 32} +{7 q^2 \over 64}\Big({53c_2\over 3}-35c_3\Big)+\bigg[m_\pi^2\Big({81 c_3-35c_2\over 2}-82c_1\Big)\nonumber \\ && +(177c_3-53c_2) {q^2\over 8}+{2m_\pi^4 \over 4m_\pi^2+q^2}(40c_1+3c_2+21c_3)-{16m_\pi^6(2c_1+3c_3)\over (4m_\pi^2+q^2)^2}\bigg] L(q) \bigg\}\,, \end{eqnarray}
\noindent$\bullet$ Spin-spin potential:
 \begin{eqnarray} V^{(0)}_\text{med}& = & {g_A^4k_n^3 \over 96\pi^4 f_\pi^6}\vec\sigma_1\!\cdot\!\vec\sigma_2  \bigg\{ \bigg[ 6m_\pi^2\Big(6c_1+c_2-{13c_3\over 2}+5c_4\Big)+ \Big(c_2-{9c_3\over 2}+{11c_4\over 3}\Big) q^2 \bigg]\ln{m_\pi \over \lambda} \nonumber \\ &&  +\Big(9c_1+{c_2\over 2}-{33c_3\over 4}+{35c_4\over 6}\Big) m_\pi^2-\Big({c_2\over 6}+{c_4\over 9}\Big) q^2 +\bigg[\Big(24c_1+2c_2-11c_3+{26c_4\over 3}\Big)
 m_\pi^2\nonumber \\ &&  +\Big(c_2-{9c_3\over 2}+{11c_4\over 3}\Big) q^2 +{4m_\pi^4 \over 4m_\pi^2+q^2}(3c_3-6c_1-2c_4)\bigg] L(q) \bigg\}\,, \end{eqnarray}
\noindent $\bullet$ Tensor potential:
 \begin{eqnarray} V^{(0)}_\text{med}& = & {g_A^4k_n^3 \over 576\pi^4 f_\pi^6}\vec\sigma_1\!\cdot\!\vec q\, \vec\sigma_2\!\cdot\!\vec q\,  \bigg\{ \Big({53c_3\over 2}-5c_2 -21c_4\Big)\ln{m_\pi \over \lambda} +{367 c_3 \over 48} - {131c_2\over 120}-{233 c_4\over 240} \nonumber \\ &&  +{4m_\pi^2 \over q^2}(18c_1+2c_2 -7c_3+6c_4) +{3m_\pi^2 \over 4m_\pi^2+q^2}(3c_3-6c_1-2c_4) \nonumber \\ && +\bigg[{53c_3 \over 2}-5c_2-21c_4  +{4m_\pi^2 \over q^2}(7c_3-18c_1-2c_2 -6c_4) \nonumber \\ && +  {3m_\pi^2 \over 4m_\pi^2+q^2}(2c_2-13c_3+10c_4)+  {12m_\pi^4 \over (4m_\pi^2+q^2)^2}(3c_3-6c_1-2c_4)\bigg] L(q) \bigg\}\,. \end{eqnarray}
\section{Subleading three-nucleon contact potential}
The subleading three-nucleon contact potential (appearing at N$^4$LO in the chiral counting) has been reexamined recently in ref.\cite{girlanda}. Its corrected expression, given in eq.(15) of ref.\cite{girlanda}, depends quadratically on momenta and it involves 13 parameters, called  $E_1,\dots, E_{13}$. Specifying to the case of three neutrons and closing two neutron-lines in all possible ways to a loop, one obtains the following contribution to the in-medium nn-potential:  
\begin{eqnarray} {V_\text{med}\over \rho_n}&=&
2(E_1+E_2)\Big( {6k_n^2\over 5}+2p^2-q^2\Big)+ 2(E_3+E_4)(\vec \sigma_1\!\cdot \!\vec\sigma_2+3)\Big({3k_n^2\over 5}+p^2\Big) \nonumber \\ && +\,3(E_5+E_6)\Big[\vec \sigma_1\!\cdot\!\vec p\, \vec \sigma_2\!\cdot \!\vec p+\vec \sigma_1\! \cdot \! \vec p\,' \vec \sigma_2\!\cdot \!\vec p\,'-
{2p^2\over 3}\vec \sigma_1\!\cdot \!\vec \sigma_2\Big] + (E_7+E_8) i (\vec \sigma_1\!+\!\vec \sigma_2)\!\cdot\!(\vec q\!\times\! \vec p\,) \nonumber \\ && +(E_9+ E_{10}) \Big[q^2-2p^2-{6k_n^2\over 5}-i (\vec \sigma_1\!+\!\vec \sigma_2)\!\cdot\!(\vec q\!\times\! \vec p\,) \Big] \nonumber \\ &&+(E_{11}+ E_{12} +E_{13}) \Big[q^2-2p^2-{6k_n^2\over 5}+i (\vec \sigma_1\!+\!\vec \sigma_2)\!\cdot\!(\vec q\!\times\! \vec p\,) \Big] \,,\end{eqnarray}
which depends linearly on the neutron density $\rho_n= k_n^3/3\pi^2$ and quadratically on the momenta 
$\vec p, \vec q, k_n$. The above expression for $V_{\rm med}$ produces non-vanishing matrix-elements only in   S- and P-wave states:
\begin{eqnarray} 
&& \langle ^1S_0|V_\text{med}|^1S_0\rangle ={ 6 \rho_nk_n^2\over 5}(2E_1+2E_2-E_9-E_{10}-E_{11}-E_{12}-E_{13} ) = {9\rho_n k_n^2\over 5} E_{c}\,,  \nonumber \\ &&  \langle ^3P_0|V_\text{med}|^3P_0\rangle ={2\over 3} \rho_n p^2(2E_1\!+\!2E_2\!+\!2E_7\!+\!2E_8\!-\!3E_9\!-\!3E_{10}\!+\!E_{11}\!+\!E_{12}\!+\!E_{13})= \rho_n p^2(E_{c}-2 E_{o})\,, \nonumber \\ &&  \langle ^3P_1|V_\text{med}|^3P_1\rangle ={2\over 3} \rho_n p^2(2E_1+2E_2+E_7+E_8-2E_9-2E_{10})= \rho_n p^2(E_{c}- E_{o})\,,\nonumber \\ && \langle ^3P_2|V_\text{med}|^3P_2\rangle ={2\over 3} \rho_n p^2(2E_1+2E_2-E_7-E_8-2E_{11}-2E_{12}-2E_{13})= \rho_n p^2(E_{c}+E_{o})\,,\end{eqnarray}
where the parts $\sim (E_3+E_4)$ and  $\sim (E_5+E_6)$ in eq.(60) are effectively zero and only a central parameter $E_c$ and a spin-orbit parameter $E_o$ are independent.
As a remarkable feature one observes that there is no term proportional to $\rho_n p^2$ in the $S$-wave matrix element $\langle ^1S_0|V_\text{med}|^1S_0\rangle$. In order to establish that this restriction is consistent with the renormalizability condition, one exploits $\vec\sigma_1=-\vec\sigma_2$ in the spin-singlet state and uses the angular average $\langle q^2\rangle = 2p^2$ of the squared momentum-transfer. By following in subsections 6.1 and 6.2 all terms that contribute proportional to  $c_{2,3}k_n^3 p^2/(64\pi^4f_\pi^6) \ln(m_\pi/\lambda)$ or $c_{2,3,4}g_A^2k_n^3 p^2/(96\pi^4f_\pi^6) \ln(m_\pi/\lambda)$ to the $S$-wave matrix element, one verifies that these indeed sum up to zero. The same exact cancelation occurs also for the terms proportional to $c_{1,2,3}k_n^3 m_\pi^2/(64\pi^4f_\pi^6) \ln(m_\pi/\lambda)$ or $c_{1,2,3,4}g_A^2k_n^3 m_\pi^2/(96\pi^4f_\pi^6) \ln(m_\pi/\lambda)$.

From the corrected version of the subleading 3N-contact potential given in eq.(15) of ref.\cite{girlanda} the following in-medium NN-potential in isospin-symmetric nuclear matter of density $\rho= 2k_f^3/3\pi^2$ is derived:
\begin{eqnarray} {V_\text {med}\over \rho}&\!\!=\!\!&
E_1\Big( {6\over 5}k_f^2+2p^2-3q^2\Big)+E_2\Big[ (\vec \tau_1\!\cdot \!\vec
\tau_2+3)\Big({3\over 5}k_f^2+p^2\Big) -\vec \tau_1\!\cdot \!\vec \tau_2\, q^2 \Big]
\nonumber \\ && + E_3\Big[( \vec \sigma_1\!\cdot \!\vec\sigma_2+3)\Big({3\over 5}k_f^2
 +p^2\Big) -\vec \sigma_1\!\cdot \!\vec \sigma_2\, q^2 \Big]+E_4\Big[(\vec \tau_1\!
\cdot \!\vec \tau_2\, \vec\sigma_1\!\cdot \!\vec\sigma_2+9)\Big({3\over 5}k_f^2+p^2\Big)
-\vec \tau_1\!\cdot \!\vec \tau_2\,\vec \sigma_1\!\cdot \!\vec \sigma_2\, q^2 \Big]
\nonumber \\ &&   +(E_5+\vec \tau_1\!\cdot \!\vec \tau_2 E_6)\Big[ \vec \sigma_1\!\cdot \!\vec 
\sigma_2(q^2-p^2)-3 \vec \sigma_1\!\cdot \!\vec q\, \vec \sigma_2\!\cdot \!\vec q+{3\over 2}
(\vec \sigma_1\! \cdot \! \vec p\, \vec \sigma_2\!\cdot \!\vec p+\vec \sigma_1\! \cdot \! \vec p\,' 
\vec \sigma_2\!\cdot \!\vec p\,')\Big]\nonumber\\ && + (7E_7-9E_8) {i \over 4} (\vec \sigma_1\!+\!\vec \sigma_2)\!\cdot\!(\vec q\!\times\! \vec p\,)+E_9 \Big[  \vec \sigma_1\!\cdot \!\vec q\, \vec \sigma_2
\!\cdot \!\vec q+{q^2\over 2}-p^2-{3\over 5} k_f^2 -{i \over 2}(\vec \sigma_1\!+\!\vec \sigma_2)\!\cdot\!(\vec q\!\times\! \vec p\,)\Big]\nonumber\\ &&  + E_{10} \Big[ \vec \tau_1\!\cdot \!\vec \tau_2\, \vec \sigma_1\!\cdot \!\vec q\, \vec \sigma_2\!\cdot \!\vec q+{3\over 2}q^2-3p^2-{9\over 5} k_f^2-{3i \over 2}
(\vec \sigma_1\!+\!\vec \sigma_2)\!\cdot\!(\vec q\!\times\! \vec p\,) \Big] \nonumber\\ &&  + E_{11} \Big[  \vec \sigma_1\!\cdot \!\vec q\, \vec \sigma_2\!\cdot \!\vec q+{q^2\over 2}-p^2-{3\over 5} k_f^2+{i \over 2}
(\vec \sigma_1\!+\!\vec \sigma_2)\!\cdot\!(\vec q\!\times\! \vec p\,) \Big] \nonumber\\ && + E_{12} \Big[ \vec \tau_1\!\cdot \!\vec \tau_2\, \vec \sigma_1\!\cdot \!\vec q\, \vec \sigma_2\!\cdot \!\vec q+{3\over 2}q^2-3p^2-{9\over 5} k_f^2+{3i \over 2}(\vec \sigma_1\!+\!\vec \sigma_2)\!\cdot\!(\vec q\!\times\! \vec p\,) \Big] \nonumber\\ &&  + {E_{13}\over 2}\Big\{\vec \tau_1\!\cdot \!\vec \tau_2  \Big[ q^2-2p^2-{6\over 5} k_f^2- \vec \sigma_1\!\cdot \!\vec q\, \vec \sigma_2\!\cdot \!\vec q+i(\vec \sigma_1\!+\!\vec \sigma_2)\!\cdot\!(\vec q\!\times\! \vec p\,) \Big]-3 \vec \sigma_1\!\cdot \!\vec q\, \vec \sigma_2\!\cdot \!\vec q\,\Big\}\,,\end{eqnarray}
which should replace the (incomplete) expression written in eq.(49) of ref.\cite{vmedshort}.


\begin{thebibliography}{99}
\bibitem{3Nlong} V. Bernard, E. Epelbaum, H. Krebs, and Ulf-G. Mei{\ss}ner, {\it 
  Phys. Rev.} {\bf C77}, 064004 (2008).\vs
\bibitem{3Nshort} V. Bernard, E. Epelbaum, H. Krebs, and Ulf-G. Mei{\ss}ner, {\it 
  Phys. Rev.} {\bf C84}, 054001 (2011).\vs 
\bibitem{twopi4} H. Krebs, A. Gasparyan, and E. Epelbaum, {\it Phys. Rev.}
  {\bf C85}, 054006 (2012).\vs 
 \bibitem{midrange4} H. Krebs, A. Gasparyan, and E. Epelbaum, {\it Phys. Rev.}
  {\bf C87}, 054007 (2013).\vs 
  \bibitem{twopidelta} H. Krebs, A.M. Gasparyan, and E. Epelbaum, {\it Phys. Rev.}
  {\bf C98}, 014003 (2018).\vs
\bibitem{holt} J.W. Holt, N. Kaiser, and W. Weise, {\it Phys. Rev.} {\bf C81},
  024002 (2010).\vs
\bibitem{vmedshort} N. Kaiser and V. Niessner, {\it Phys. Rev.} {\bf C98},
  054002 (2018).\vs
\bibitem{vmedlong} N. Kaiser and B. Singh, {\it Phys. Rev.} {\bf C100},  014002 (2019).\vs
\bibitem{subsubvmed} N. Kaiser, {\it Phys. Rev.} {\bf C101},  014001 (2020).\vs
  \bibitem{normalorder} C. Drischler, K. Hebeler, and A. Schwenk, {\it Phys. Rev.} {\bf C93}, 054314 (2016);  {\it Phys. Rev.  Lett.} {\bf 122}, 042501 (2019); and references therein.\vs
  \bibitem{achim1}  C. Drischler, A. Carbone, K. Hebeler, and A. Schwenk, {\it
  Phys. Rev.} {\bf C94}, 054307 (2016).\vs
 \bibitem{treuer}  L. Treuer, Bachelor thesis, TU Munich, nucl-th-2009.11104.\vs
 \bibitem{girlanda} L. Girlanda, A. Kievsky, and M. Viviani, {\it Phys. Rev.} {\bf C102}, 019903(E) (2020).\vs
\end{thebibliography}
\end{document}